\documentclass{PoS}

\title{Geometrical Origin of CP Violation and CKM and MNS Matrices in $SU(5) \times T^{\prime}$}

\ShortTitle{Geometrical Origin of CP Violation ... in $SU(5) \times T^{\prime}$}

\author{Mu-Chun Chen\\
        Department of Physics \& Astronomy, University of California, Irvine, CA 92697-4575, U.S.A.\\
        E-mail: \email{muchunc@uci.edu}}

\author{\speaker{K.T. Mahanthappa} \\ 
        Department of Physics, University of Colorado, Boulder, CO 80309-0390, U.S.A.\\
        E-mail: \email{ktm@colorado.edu}}

\abstract{We propose the complex group theoretical Clebsch-Gordon coefficients as a novel origin of CP violation. This is manifest in our model based on SUSY SU(5)  combined with the double tetrahedral group, $T^{\prime}$, as a family symmetry. Due to the presence of the doublet representations in $T^{\prime}$, there exist complex CG coefficients, leading to explicit CP violation in the model, while the Yukawa couplings and the vacuum expectation values of the scalar fields remain real. The tri-bimaximal neutrino mixing matrix arises from the CG coefficients of $T^{\prime}$. In addition to the prediction for 
$\theta_{13} \sim (1/ (3\sqrt{2}) \theta_{c}$, 
the model gives rise to a sum rule, 
$\tan^{2} \theta_{\odot} \sim  \tan^{2} \theta_{\odot \; ,\mbox{\tiny TBM}} + (1/2) \theta_{c} \cos \delta_{\ell}$, 
which is a consequence of the Georgi-Jarlskog relations in the charged fermion sector. The predicted leptonic Dirac CP phase, $\delta_{\ell}$, gives the correct value of the solar mixing angle, and the predicted CP violation measures in the quark sector are consistent with the current experimental data. Since the Dirac CP phase is the only non-vanishing phase predicted in the lepton sector, there is a connection between leptogenesis and low energy leptonic CP violating processes in our model.
}

\FullConference{35th International Conference of High Energy Physics - ICHEP2010,\\
		July 22-28, 2010\\
		Paris France}

\begin{document}

\section{Introduction}

The origin of the cosmological matter antimatter asymmetry in the universe is one of the fundamental questions that still remain to be answered.  The observation of neutrino oscillation opens up the possibility of generating the baryonic asymmetry through leptogenesis. The success of leptogenesis crucially depends on the existence of CP violating phases in the Pontecorvo-Maki-Nakagawa-Sakata (PMNS) matrix that describes the neutrino mixing~\cite{Chen:2007fv}.

We propose~\cite{Chen:2009gf} the complex Clebsch-Gordon (CG) coefficients as a new origin of CP violation. Such complex CG coefficients exist~\cite{cg} in the double tetrahedral group, $T^{\prime}$~\cite{dt,Chen:2007afa,Chen:2009gf}. (The symmetry $T^{\prime}$ has also been utilized as a solution to the FCNC problem in the Randall-Sundrum model. See~\cite{Chen:2009gy}.) In this scenario, CP violation occurs explicitly from the CG coefficients of the $T^{\prime}$ group theory, while the Yukawa coupling constants and the VEVs of the scalar fields remain real. As a result, the amount of CP violation in our model is determined entirely by the group theory, unlike in the usual scenarios. It gives the right amount of CP violation in the quark sector and in the lepton sector, through leptogenesis, leads to the right amount of the matter antimatter asymmetry.

\section{The Model}

Our model~\cite{Chen:2009gf,Chen:2007afa}, which is based on SUSY $SU(5)$ model combined with the $T^{\prime}$ family symmetry,  simultaneously gives rise to the tri-bimaximal neutrino mixing and realistic CKM quark mixing. For the field content of our model, see~\cite{Chen:2009gf,Chen:2007afa}.  The discrete symmetry, $Z_{12}\times Z_{12}^{\prime}$, of our model allow the lighter generation masses to arise only at higher mass dimensionality, and thus provides a dynamical origin of the mass hierarchy. The superpotential of the Yukawa sector of the model is given by,  
$\mathcal{W}_{\mbox{\tiny Yuk}} =  \mathcal{W}_{TT} + \mathcal{W}_{TF} + \mathcal{W}_{\nu}$, where  
\begin{eqnarray}
\mathcal{W}_{TT} & = & y_{t} H_{5} T_{3} T_{3} + \frac{1}{\Lambda^{2}}  H_{5} \biggl[ y_{ts} T_{3} T_{a} \psi \zeta + y_{c} T_{a} T_{b} \phi^{2} \biggr] + \frac{1}{\Lambda^{3}} y_{u} H_{5} T_{a} T_{b} \phi^{\prime 3} \quad \label{eq:Ltt} \\ 
\mathcal{W}_{TF} & = &  \frac{1}{\Lambda^{2}} y_{b} H_{\overline{5}}^{\prime} \overline{F} T_{3} \phi \zeta + \frac{1}{\Lambda^{3}} \biggl[ y_{s} \Delta_{45} \overline{F} T_{a} \phi \psi \zeta^{\prime}  + y_{d} H_{\overline{5}^{\prime}} \overline{F} T_{a} \phi^{2} \psi^{\prime} \biggr]  \quad   \label{eq:Ltf} \\
\mathcal{W}_{\nu} & = & \lambda_{1} NNS +  \frac{1}{\Lambda^{3}} \biggl[ H_{5}  \overline{F} N \zeta \zeta^{\prime} \biggl( \lambda_{2} \xi + \lambda_{3} \eta\biggr) \biggr] \quad
\label{eq:Lff}
\end{eqnarray}
which is invariant under $SU(5) \times T^{\prime}$ and it is CP non-invariant. 
Note that all Yukawa coupling constants, $y_{x}$, are real parameters since their phases can be absorbed by redefinition of the Higgs and flavon fields. 
The vacuum expectation values of the $T^{\prime}$ flavon fields can be found in Ref.~\cite{Chen:2009gf}. Note that  all the expectation values are real. 

The matrices $M_{u}$, $M_{d}$ and $M_{e}$, upon the breaking of $T^{\prime}$ and the electroweak symmetry, are given in terms of seven parameters by~\cite{Chen:2009gf,Chen:2007afa}
\begin{eqnarray}
\frac{M_{u}}{y_{t}v_{u}} =  \left( \begin{array}{ccc}
i \phi^{\prime 3}_{0}  & (\frac{1-i}{2}) \phi_{0}^{\prime 3} & 0 \\
(\frac{1-i}{2})  \phi_{0}^{\prime 3}  & \phi_{0}^{\prime 3} + (1 - \frac{i}{2}) \phi_{0}^{2} & y^{\prime} \psi_{0} \zeta_{0} \\
0 & y^{\prime} \psi_{0} \zeta_{0} & 1
\end{array} \right), 
\frac{M_{d} , \; M_{e}^{T}}{ y_{d} v_{d} \phi_{0} }   =  \left( \begin{array}{ccc}
0 & (1+i) \phi_{0} \psi^{\prime}_{0} & 0 \\
-(1-i) \phi_{0} \psi^{\prime}_{0} & (1,-3) \psi_{0} N_{0} & 0 \\
\phi_{0} \psi^{\prime}_{0} & \phi_{0} \psi^{\prime}_{0} & \zeta_{0} 
\end{array}\right)\quad .
\end{eqnarray}
The $SU(5)$ relation, $M_{d} = M_{e}^{T}$, is manifest in the above equations, except for the factor of $-3$ in the (22) entry of $M_{e}$, due to the $SU(5)$ CG coefficient through the coupling to $\Delta_{45}$. In addition to this $-3$ factor, the Georgi-Jarlskog (GJ) relations also require $M_{e,d}$ being non-diagonal, leading to corrections to the TBM pattern~\cite{Chen:2009gf,Chen:2007afa}.  Note that the complex coefficients in the above mass matrices arise {\it entirely} from the CG coefficients of $T^{\prime}$. 

The interactions in $\mathcal{W}_{\nu}$ lead to the following neutrino mass matrix, 
\begin{equation}
M_{RR} = \left( \begin{array}{ccc}
1 & 0 & 0 \\
0 & 0 & 1 \\
0 & 1 & 0 
\end{array}\right) s_{0} \Lambda \; , 
M_{D} = \left( \begin{array}{ccc}
2\xi_{0} + \eta_{0} & -\xi_{0} & -\xi_{0} \\
-\xi_{0} & 2\xi_{0} & -\xi_{0} + \eta_{0} \\
-\xi_{0} & -\xi_{0}+\eta_{0} & 2\xi_{0} 
\end{array}\right) \zeta_{0} \zeta^{\prime}_{0} v_{u} \; .
\end{equation}
As these interactions involve only the triplet representations of $T^{\prime}$, all relevant CG coefficients are real, leading to a real neutrino Majorana mass matrix. The resulting effective neutrino mass matrix $M_{D} M_{RR}^{-1} M_{D}^{T}$  has the special property that it is form diagonalizable by the TBM mixing matrix,
$U_{\mbox{\tiny TBM}}^{T} M_{\nu} U_{\mbox{\tiny TBM}}  =    \mbox{diag}((u_{0} + 3 \xi_{0})^{2}, u_{0}^{2}, -(-u_{0}+3\xi_{0})^{2}) \frac{v_{u}^{2}}{M_{X}}$ 
 $\equiv   \mbox{diag} (m_{1}, m_{2}, m_{3})$.
While the neutrino mass matrix is real, the complex charged lepton mass matrix leads to a complex 
$V_{\mbox{\tiny PMNS}} = V_{e, L}^{\dagger} U_{\mbox{\tiny TBM}}$.

\section{Numerical Predictions}

The predicted charged fermion mass matrices in our model are parametrized in terms of 7 parameters~\cite{Chen:2009gf,Chen:2007afa},
\begin{eqnarray}
\frac{M_{u}}{y_{t} v_{u}} = \left( \begin{array}{ccccc}
i g & ~~ &  \frac{1-i}{2}  g & ~~ & 0\\
\frac{1-i}{2} g & & g + (1-\frac{i}{2}) h  & & k\\
0 & & k & & 1
\end{array}\right)  , \quad 
\frac{M_{d}, \; M_{e}^{T}}{y_{b} v_{d} \phi_{0}\zeta_{0}}  =  \left( \begin{array}{ccccc}
0 & ~~ & (1+i) b & ~~ & 0\\
-(1-i) b & & (1,-3) c & & 0\\
b & &b & & 1
\end{array}\right)  \; .
\end{eqnarray}
With $b \equiv \phi_{0} \psi^{\prime}_{0} /\zeta_{0} = 0.00304$, $c\equiv \psi_{0}N_{0}/\zeta_{0}=-0.0172$,  $k \equiv y^{\prime}\psi_{0}\zeta_{0}=-0.0266$, $h\equiv \phi_{0}^{2}=0.00426$ and $g \equiv \phi_{0}^{\prime 3}= 1.45\times 10^{-5}$, the following mass ratios are obtained, 
$m_{d}: m_{s} : m_{b} \simeq \theta_{c}^{\scriptscriptstyle 4.7} : \theta_{c}^{\scriptscriptstyle 2.7} : 1$, 
$m_{u} : m_{c} : m_{t} \simeq  \theta_{c}^{\scriptscriptstyle 7.5} : \theta_{c}^{\scriptscriptstyle 3.7} : 1$, 
with $\theta_{c} \simeq \sqrt{m_{d}/m_{s}} \simeq 0.225$. We have also taken $y_{t} = 1.25$ and $y_{b}\phi_{0} \zeta_{0} \simeq m_{b}/m_{t} \simeq 0.011$ and included the renormalization group corrections. As a result  of the GJ relations, realistic charged lepton masses are obtained. 
Making use of these parameters, the complex CKM matrix (in standard form) is,
\begin{eqnarray}
\left( \begin{array}{ccc}
0.974 & 0.227  & 0.00412e^{-i45.6^{o}} \\
-0.227 - 0.000164 e^{i45.6^{o}} & 0.974 - 0.0000384 e^{i45.6^{o}} & 0.0411 \\
0.00932 - 0.00401 e^{i45.6^{o}} & -0.0400 - 0.000935 e^{i45.6^{o}} & 1
\end{array}\right). 
\end{eqnarray}
Values for all $|V_{\mbox{\tiny CKM}}|$ elements are consistent with current experimental values except for $|V_{td}|$, the experimental determination of which has large hadronic uncertainty.  
The predictions of our model for the angles in the unitarity triangle are, 
$\beta = 23.6^{o}$, $\alpha = 110^{o}$, and $\gamma = \delta_{q} = 45.6^{o}$, 
 and they agree with the direct measurements within $1\sigma$ of BaBar and $2\sigma$ of Belle.  Our predictions for the Wolfenstein paramteres, $\lambda=0.227$, $A=0.798$, $\overline{\rho} = 0.299$ and $\overline{\eta}=0.306$, are very close to the global fit values except for $\overline{\rho}$. Our prediction for the Jarlskog invariant, $J  \equiv  \mbox{Im} (V_{ud} V_{cb} V_{ub}^{\ast} V_{cd}^{\ast}) = 2.69 \times 10^{-5}$, in the quark sector also agrees with the current global fit value.

As a result of the GJ relations, our model predicts  $\tan^{2} \theta_{\odot} \simeq \tan^{2} \theta_{\odot,\mbox{\tiny TBM}} + \frac{1}{2} \theta_{c} \cos\delta_{\ell}$, with $\delta_{\ell}$ being the leptonic Dirac CP phase in the standard parametrization.  
In addition, our model predicts $\theta_{13} \sim \theta_{c}/3\sqrt{2}$. Numerically, the diagonalization matrix for the charged lepton mass matrix combined with $U_{TBM}$ gives the PMNS matrix (in standard form), 
\begin{equation}
\left( \begin{array}{ccc}
0.838  & 0.542 & 0.0583 e^{-i227^{o}}   \\
-0.385  - 0.0345 e^{i227^{o}} & 0.594 - 0.0224 e^{i227^{o}} & 0.705   \\
0.384 - 0.0346 e^{i227^{o}} & -0.592 - 0.0224 e^{i227^{o}} & 0.707
\end{array}\right) \; ,
\end{equation}
which gives $\sin^{2}\theta_{\mathrm{atm}} = 1$, $\tan^{2}\theta_{\odot} = 0.420$ and $|U_{e3}| = 0.0583$. The VEV's, $\xi_{0} = -0.0791$ and $\eta_{0} = 0.1707$, $s_{0} \Lambda= 10^{12}$ GeV, give $\Delta m_{atm}^{2} = 2.5 \times 10^{-3} \; \mbox{eV}^{2}$ and $\Delta m_{\odot}^{2} = 7.6 \times 10^{-5} \; \mbox{eV}^{2}$. The leptonic Jarlskog is predicted to be $J_{\ell} = -0.00967$, and equivalently, this gives $\delta_{\ell} = 227^{o}$.  Our model predicts $(m_{1},m_{2},m_{3}) = (0.00134,0.00882,0.0504)$ eV, with Majorana phases $\alpha_{31} = \pi$ and $\alpha_{21}$ = 0.  
 Since $\delta_{\ell}$ is the only non-vanishing leptonic CP violating phase, our model also predicts the sign of the baryonic asymmetry. 
$\delta_{\ell}$ is the only non-vanishing leptonic CP violating phase in our model and it gives rise to lepton number asymmetry, $\epsilon_{\ell} \sim 10^{-6}$. By virtue of leptogenesis, this gives the right sign and magnitude of the matter-antimatter asymmetry~\cite{CM}.

\section{Conclusion}

We propose the complex group theoretical CG coefficients as a novel origin of CP violation. This is manifest in our model based on SUSY SU(5)  combined with the double tetrahedral group, $T^{\prime}$,  as a family symmetry. Due to the presence of the doublet representations in $T^{\prime}$, there exist complex CG coefficients, leading to explicit CP violation in the model. 
The predicted CP violation measures in the quark sector are consistent with the current experimental data. The leptonic Dirac CP violating phase is predicted to be $\delta_{\ell} \sim 227^{o}$, which is close to current SuperK best fit value of $220^{o}$ and it can give the cosmological matter asymmetry.

\section*{Acknowledgments}
The work of KTM was supported, in part, by the Department of Energy under Grant No. DE-FG02-04ER41290.  The work of M-CC was supported, in part, by the National Science Foundation under Grant No. PHY-0709742 and PHY-0970173.

\end{document}